\documentstyle[sprocl]{article}

\def\Journal#1#2#3#4{{#1} {\bf #2}, #3 (#4)}

\def\PRD{{\em Phys. Rev.} D}

\def\be{\begin{equation}}
\def\ee{\end{equation}}
\def\bea{\begin{eqnarray}}
\def\eea{\end{eqnarray}}

\begin{document}

\title{EXACT COSMOLOGICAL SOLUTIONS OF NONLINEAR $F(R)$-GRAVITY}

\author{H.-J. SCHMIDT}

\address{Universi\"at Potsdam\\Institut f\"ur Mathematik\\Am
Neuen
Palais 10\\D-14415~Potsdam, Germany\\E-mail:
 hjschmi@rz.uni-potsdam.de}

\maketitle\abstracts{We report on the cited papers
refs. 1 - 18 from the
following points of view: What do we exactly know about solutions
when no exact solution (in the sense of ``solution in closed
form'') is
available? In which sense do 
these solutions possess a singularity? 
In which cases do
 conformal relations and/or dimensional reductions
simplify the deduction? Furthermore,
 we outline some open questions
 worth of being studied in future research.}

\bigskip

\section{Singularity Theorem}

In ref. 1  the following simple type of singularity 
theorems was discussed: 
The coordinates $t, \ x, \ y, \ z$ shall cover all 
the reals, and $a(t)$ shall be an arbitrary strictly 
positive monotonously increasing smooth function defined 
for all
 real values $t$, where ``smooth'' denotes 
``$C^{\infty}$-differentiable''. Then it holds

\noindent 
{\bf Lemma 1}: The Riemannian space defined by 
$$ds^2= dt^2+a^2(t)(dx^2+dy^2+dz^2)$$
 is geodesically complete.
 
This fact is well-known and easy to prove; 
however, on the other hand it holds for the same
class of functions $a(t)$ 

\noindent 
{\bf Lemma 2}: The Pseudoriemannian space--time defined by 
\begin{equation}
ds^2= dt^2-a^2(t)(dx^2+dy^2+dz^2)
\end{equation}
 is light--like geodesically complete iff 
\begin{equation}
\int_{- \infty}^0 \ a(t) \ dt \ \ = \ \ \infty
\end{equation}

As usual, ``iff'' denotes ``if and only if''.
The proof is straightforwardly done by considering light--like
 geodesics in the $x-t$-plane. So, one directly 
concludes that the statements do not depend on
the number $n$ of spatial dimensions, (only $n>0$ is
used), and the formulation for $n=3$ was
chosen as the most interesting case.

Moreover,  
 lemma 2 remains valid if we replace 
``light--like geodesically complete'' by 
``light--like and time--like geodesically complete''.

In ref. 1 I wrote: ``Lemma 2 seems to be unpublished
 up to now.'' Now I can add two earlier references
\cite{2}, \cite{3}
which presented related statements:  
Borde, Vilenkin \cite{2}, besides considering more general
inhomogeneous space-times, use metric (1) in conformal
time $\eta$, i.e., $dt = a(\eta)d\eta$. Then we have
instead of eq. (2) (see the appendix of \cite{2})
the following condition
\begin{equation}
\int_{\eta_{\rm min}}^\eta \ a^2(\hat \eta) \ d \hat \eta
\ \ = \ \ \infty
\end{equation}
Up to this time-reparametrization their result coincides 
with our lemma 2. 

Romero, Sanchez \cite{3}, see also Sanchez \cite{4},
have discussed a quite general class of warped product
space--times and the conditions of geodesic completeness
in them, and our lemma 2 can be found after some tedious
reformulations and specializations from Theorem 3.9. 
of \cite{3}.

Further recent results on singularity theorems are
reviewed in Senovilla \cite{5}. That review and most of
the research concentrated mainly on the 4--dimensional
Einstein theory. Probably, the majority of arguments 
can be taken over to the higher--dimensional Einstein
equation without change, but this has not yet worked
out up to now. And  singularity theorems for
$F(R)$-gravity are known up to now  for very
special cases only.

\bigskip

\section{Rigorous Solutions}

The most frequently used $F(R)$-Lagrangian is 
\be 
L = \left( \frac{R}{2} \ - \  \frac{l^2}{12} \, R^2 \right) 
\ \sqrt{-g}  \qquad {\rm where} \  l > 0.
\ee
\noindent 
Here we discuss the non--tachyonic case only. 
From Lagrangian (4) one gets a fourth-order field equation. 
Only very few closed-form solutions (``exact solutions'')
are known. However, for the class of spatially flat
Friedmann models, eq. (1), the set of solutions is
qualitatively 
completely described (but not in closed form) 
in M\"uller, Schmidt \cite{6}.
We call them ``rigorous solutions''. 

One of them, often called ``Starobinsky inflation'', 
can be approximated by eq. (1) with 
\begin{equation} 
a(t) \ = \ \exp( - \frac{t^2}{12 l^2} )
\end{equation}
This  approximation is valid in the region
$t \ll - l$.
However, this solution does not fulfil the
condition eq. (2). Therefore, by lemma 2, 
Starobinsky inflation does not 
represent a light--like geodesically complete cosmological 
model as has been frequently stated in the literature.

To prevent a further misinterpretation let me
reformulate this result as follows: 
Inspite of the fact that the Starobinsky 
 model is regular (in the sense that $a(t) > 0$ for arbitrary
values of synchronized time $t$), every past--directed 
light--like geodesic terminates in a curvature
singularity (i.e.,  $\vert R \vert \longrightarrow \infty$)
 at a finite value of its affine parameter.
Therefore, the model is not only geodesically
incomplete in the coordinates chosen, but it also
fails to be a subspace of a complete space--time.

In contrast to this one can say:  
 Eq. (1) with $a(t) = \exp(H t)$, $H$ being a positive
constant, is the inflationary de Sitter space--time. 
According to lemma 2, it is also incomplete. 
However,  contrary to the Starobinsky model, 
it is a subspace of a complete space--time, namely
the de Sitter space-time represented as a closed Friedmann model.

For the Hamiltonian formulation of fourth-order gravity 
see e.g. Demaret, Querella \cite{7} and Schmidt \cite{8} and
the cited references.

The case that the conformally invariant 
term $C_{ijkl} C^{ijkl}$ is also part of 
 the gravitational 
Lagrangian has been quite often mentioned in the
literature, e.g. as the most natural one besides Einstein 
gravity. On the other hand, besides ref. 6 
only very few rigorous solutions are known for this case.

\bigskip

\section{Dimensional Reduction}

There exist several possibilites for a 
dimensional reduction of higher-dimensio\-nal space--times
with some symmetries to a corresponding lower-dimensional
space-time with additional fields; e.g.   
the 4-dimensional metrics with 2 commuting Killing vectors
 can be  reduced to a 2-dimensional metric with 2 
additional scalar fields.  \cite{9} 
Moreover, solutions of the
corresponding gravity theories go over to solutions by
this procedure. 

For  2-dimensional gravity theories the closed--form
solutions 
have been carefully studied recently, see e.g. 
Kl\"osch, Strobl \cite{10} and the references cited there. 

So, even if there might be doubts about the physical 
significance of lower- or higher-dimensional gravity
theories, one can apply the results of them (e.g.
the 
presentation and classification of exact solutions) to 
give a mathematical relation to corresponding classes
of 4--dimensional space-times.

Surprisingly, 
up to now this possibility to search 
 for new solutions of Einstein's equation in 4 dimensions 
has not been  applied very often.

\bigskip

\section{Conformal Transformation}

Besides dimensional reduction,
 the application of conformal transformations 
 represents one of 
 the most powerful methods to transform 
different theories into each other.
 In many cases this procedure 
 related theories to each other 
 which have been
originally considered to be independent ones. 
     The most often discussed question  in this context 
is ``which of the metrics is the physical one'', but 
one can, of course, use such conformal relations  
also as a simple mathematical tool to find exact solutions
without the necessity of answering this question.
     However, concerning a conformal transformation of 
 singularity theorems one must
be cautious, because typically, near a singularity in one of the
theories, the conformal factor diverges, and then the
conformally transformed metric need not be singular there.

  For transformations
 of classes of theories 
 containing $F(R)$-theories 
see e.g. \cite{11} and 
Capozziello et al. \cite{12}, and for 
 transformations of 
sixth--order
theories stemming from $R\Box R$-Lagrangains 
 see e.g. Wands \cite{13}.

\bigskip

\section{Exact Solutions}

A lot of papers claim 
to have presented new exact solutions of the Einstein equation
(in arbitrary dimension), but in many of these cases the 
authors did not carefully check whether their solution
is only a well-known old solution in unusual coordinates. 
 To simplify this check,  Mignemi, Schmidt \cite{14}
wrote down the $n$-dimensional de Sitter space--time
in several classes of unusual coordinates. 
 Maroto, Shapiro \cite{15} and
the book by  Krasinski \cite{16} represent further
sources for methods to
 characterize exact cosmological solutions.   

\medskip

Finally, let me again stress the 
 important difference  
of the Euclidean and the Lorentzian signature: 
 Smooth connected complete Riemannian spaces are 
geodesically connected. However, 
smooth connected complete space--times 
(e.g. the de Sitter space--time) need not be 
geodesically connected (see  for instance 
Sanchez \cite{17}).

The topological origin of this distinction 
 is the same as
the distinction  between lemma 1 and lemma 2: lemma 1
uses the compactness of the rotation group, whereas 
lemma 2 uses the  
noncompactness of the Lorentz group \cite{18}.

\newpage

\bigskip

\section*{Acknowledgments}
I thank A. Borde, U. Kasper,  
W. K\"uhnel, S. Mignemi, L. Querella, 
 M. Sanchez and I. Shapiro for valuable comments, and 
 DFG and HSP~III for financial support.


\section*{References}
\begin{thebibliography}{99}
\bibitem{1} H.-J. Schmidt, \Journal{\PRD}{54}{7906}{1996}
\bibitem{2} A. Borde and A. Vilenkin, 
{\it The Impossibility of Steady--State Inflation,}
 gr-qc/9403004
\bibitem{3} A. Romero and M. Sanchez, {\it Geom. Dedicat.}
{\bf 53}, 103 (1994) 
\bibitem{4} M. Sanchez, {\it Gen. Relat. Grav.} {\bf 30},
 915 (1998) 
\bibitem{5} J. Senovilla, {\it Gen. Relat. Grav.} {\bf 30},
 701 (1998) 
\bibitem{6} V. M\"uller and H.-J. Schmidt, {\it Gen. Relat.
Grav.} {\bf 17},
 769 and 971 (1985) 
\bibitem{7} J. Demaret and L. Querella, {\it 
Class. Quantum Grav.} {\bf 12}, 3085 (1995) 
\bibitem{8} H.-J. Schmidt,  gr-qc/9712097;
 {\it Grav. and Cosmol.} {\bf 3}, 266 (1997)
\bibitem{9} H.-J. Schmidt, {\it A two-dimensional representation
of four-dimensional gravitational waves,}  gr-qc/9712034; 
{\it Int. J. Mod. Phys.} D {\bf 7}, 215 (1998)
\bibitem{10}
T. Kl\"osch and T. Strobl, {\it Phys. Rev.} D {\bf  57}, 1034
(1998) 
\bibitem{11} H.-J. Schmidt,  gr-qc/9703002;
{\it Gen. Relat. Grav.} {\bf 29}, 859 (1997)
\bibitem{12} S. Capozziello, R. de Ritis and
 A. Marino, {\it Class. Quantum Grav.} {\bf 14}, 3243 (1997) 
\bibitem{13} D. Wands, {\it Class. Quantum Grav.}
 {\bf 11}, 269 (1994)
\bibitem{14} S. Mignemi and H.-J. Schmidt, gr-qc/9709070;
{\it J. Math. Phys.} {\bf 39}, 998 (1998)
\bibitem{15} A. Maroto and I. Shapiro,
{\it Phys. Lett.} B {\bf  414}, 34 (1997)
\bibitem{16} A. Krasinski, {\it Inhomogeneous 
Cosmological Models},  
  (Cambridge University Press, 1997)
\bibitem{17} M. Sanchez, {\it Gen. Relat. Grav.} {\bf 29},
 1023 (1997) 
\bibitem{18} H.-J. Schmidt, {\it Int. J. Theor. Phys.}
 {\bf 37}, 691 (1998)

\end{thebibliography}
\end{document}